# Site-specific ILC Detector Installation Plan

Karsten Buesser[1]
Thomas Schoerner[2]

Deutsches-Elektronen-Synchrotron DESY
Notkestr. 85, 22607 Hamburg, Germany

[1] karsten.buesser@desy.de
[2] thomas.schoerner@desy.de

Grant Agreement No: 645479

# E-JADE

Europe-Japan Accelerator Development Exchange Programme
Horizon 2020 / Marie Skłodowska-Curie Research and Innovation Staff Exchange (RISE)

## DELIVERABLE REPORT

## MDIPLAN

### DELIVERABLE: 22

| | |
|---|---|
| **Document identifier:** | E-Jade.Deliverable.WP3.D22.MDIPlan.v3 |
| **Due date of deliverable:** | End of month 48 (December 2018) |
| **Report release date:** | 20/12/18 |
| **Work package:** | WP3: Linear collider targeted R&D |
| **Lead beneficiary:** | DESY |
| **Document status:** | Final / Public |

**Delivery Slip**

| | Name | Partner | Date |
|---|---|---|---|
| **Authored by** | K. Buesser [ILD MDI coordinator]<br>T. Schörner-Sadenius [Project Manager] | DESY | 03/12/18 |
| **Reviewed by** | M. Stanitzki [WP3 leader] | DESY<br>DESY | 10/12/18 |
| **Approved by** | General Assembly | | 17/06/18 |

**Deliverable:**

D22: MDIPlan – Site-specific ILC detector installation plan

**Executive summary:**

Both ILC detector concepts – ILD and SiD – are very complex machines, the assembly and installation of which in the experimental cavern at the ILC interaction point will be complicated endeavours. These procedures require careful planning and logistics, taking into account numerous constraints and boundary conditions. Some of these are described in this document. Especially ILD has already invested significant effort into elaborate installation plans, which will be briefly described in this document.

However, with the ILC not being a secured project and a final interaction point not chosen, all these plans have to be considered preliminary and need to be further detailed, taking into account the concrete site-specific and project-specific conditions.

## 1. INTRODUCTION

The International Linear Collider (ILC) is an envisaged 20-30 km linear accelerator currently under scrutiny by the Japanese government for construction in the Kitakami mountain region, see Figs. 1 and 2. The ILC – once approved – will be a multinational multi-billion Euro project with a construction period of the order of 10 years.

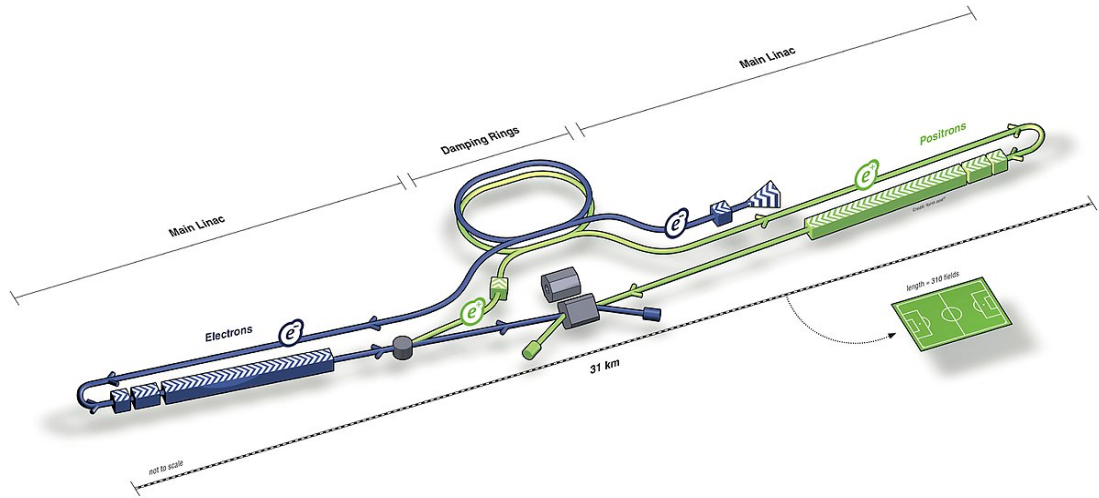

*Figure 1: Schematic overview of the ILC machine* (ILC Comms, cc-by-sa 3.0).

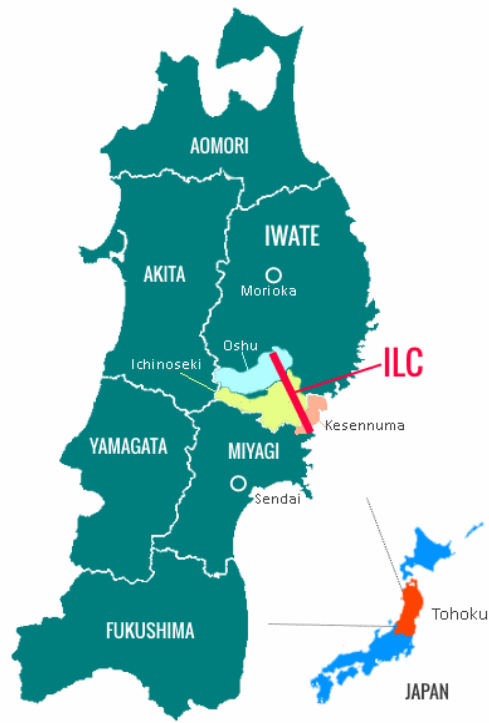

*Figure 2: The Japanese Tohoku region and the possible location of the ILC* [1].

Such a massive project poses numerous challenges (beyond the political and funding ones): logistics, transportation, construction, assembly, safety (mechanics, electricity, gas, radiation, …), services, infrastructures – to name just a few. For the accelerator – which is not the topic of this report – the largest challenge will be construction of the tunnel itself and the industrialised mass production of the modules and their timely transportation to and installation in the tunnel. This latter task requires a well-understood transportation system that connects manufacturers of components via sea, rail and road with the Kitakami site, taking into account issues like the maximum allowed load on trucks, the capacity of local harbours, the availability of storage space nearby the tunnel area, etc. These issues have also been discussed extensively – but not finally solved – at the numerous ILC infrastructure workshops in which European researchers could participate in the past few years thanks to the E-JADE funding [1], c.f. section 5. The issue of accelerator component production is also discussed in the European preparation plan [2], as is reported briefly in another recent E-JADE deliverable report [3].

Also, the construction and assembly of the detectors – according to today's view ILD and SiD in a push-pull configuration – is a difficult task, which in addition has to fit into the tight accelerator construction schedule. This issue will be discussed in greater detail in the next section.

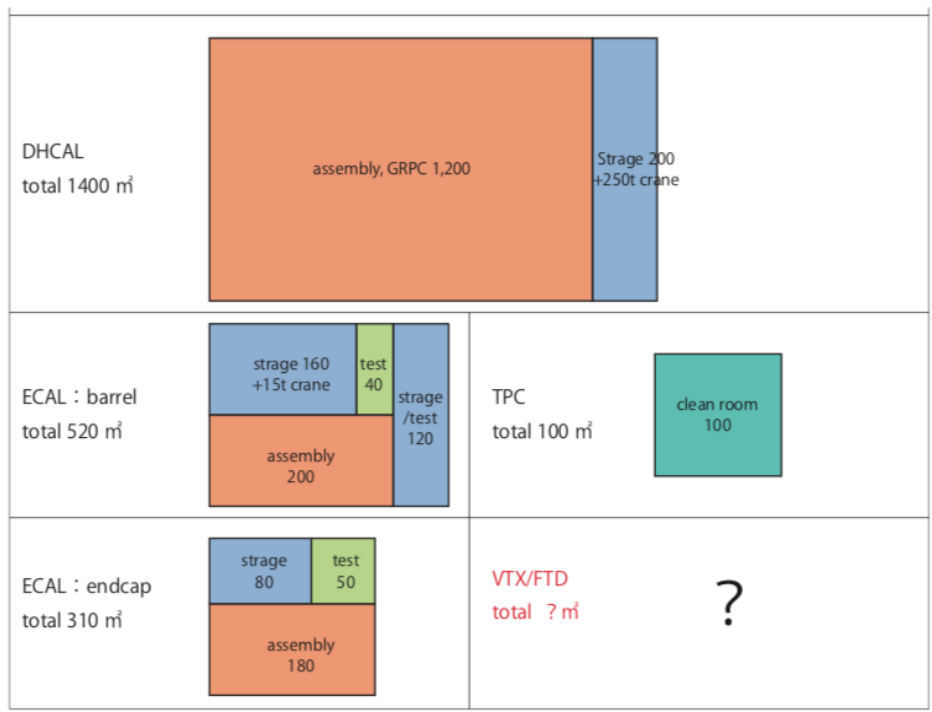

*Figure 3: ILD assembly space survey.*

## 2. ILC EXPERIMENTS CONSTRUCTION AND INSTALLATION CONSIDERATIONS

The construction and assembly of the ILC experiments ILD and SiD requires a well-prepared construction site (i.e. the later interaction region site with technical and laboratory infrastructure) and, later in the process, the finalised experimental cavern. Central to the endeavour are transportation and storage (for the delivery of many heavy parts, especially for the detector yokes and calorimeter parts), subdetector assembly areas, a large central assembly hall (AH), the detector hall (DH) for final installation, and the necessary services (power, gas, cranes, lab space, workshops, etc.

In order to reasonably plan all these facilities, several surveys have been performed among ILD and SiD and their respective sub-components, with the aim of finding out about the individual needs concerning space and services, and the individual timelines. Of course, all these plans are still very preliminary and subject to change because of changing boundary conditions once the project is approved. Figure 3 shows, as an example, the result of a survey for assembly, storage and test space for individual ILD subdetectors performed in 2016. Since then, the survey has been updated, and rather robust numbers have been obtained for both experiments.

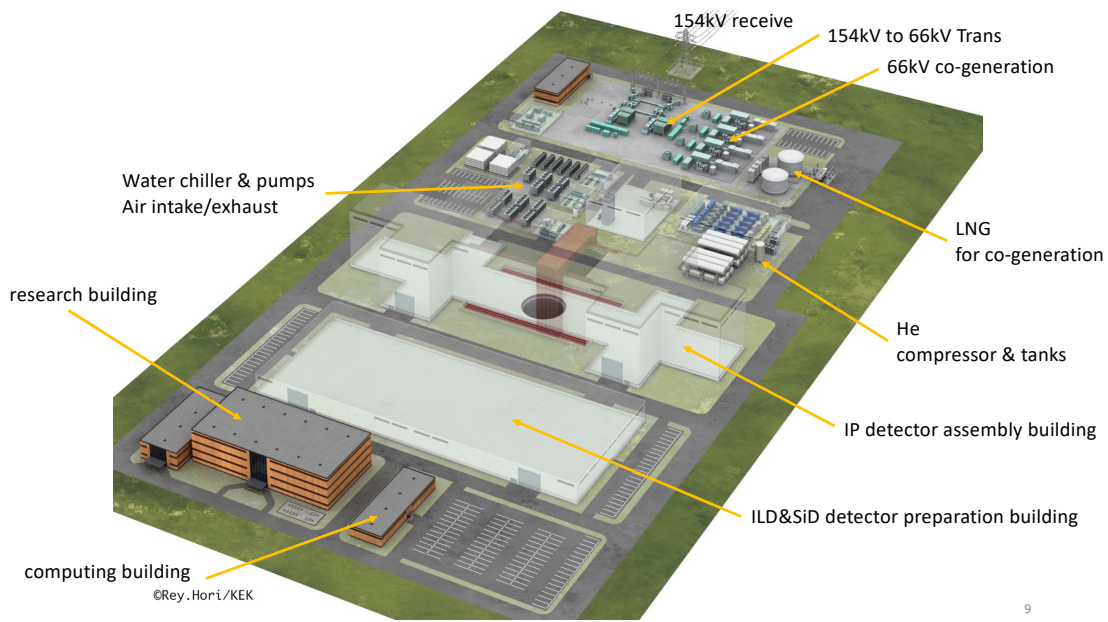

*Figure 4: Proposal for the layout of the interaction region campus in the Kitakami area. Since the* ... *ty* [4].

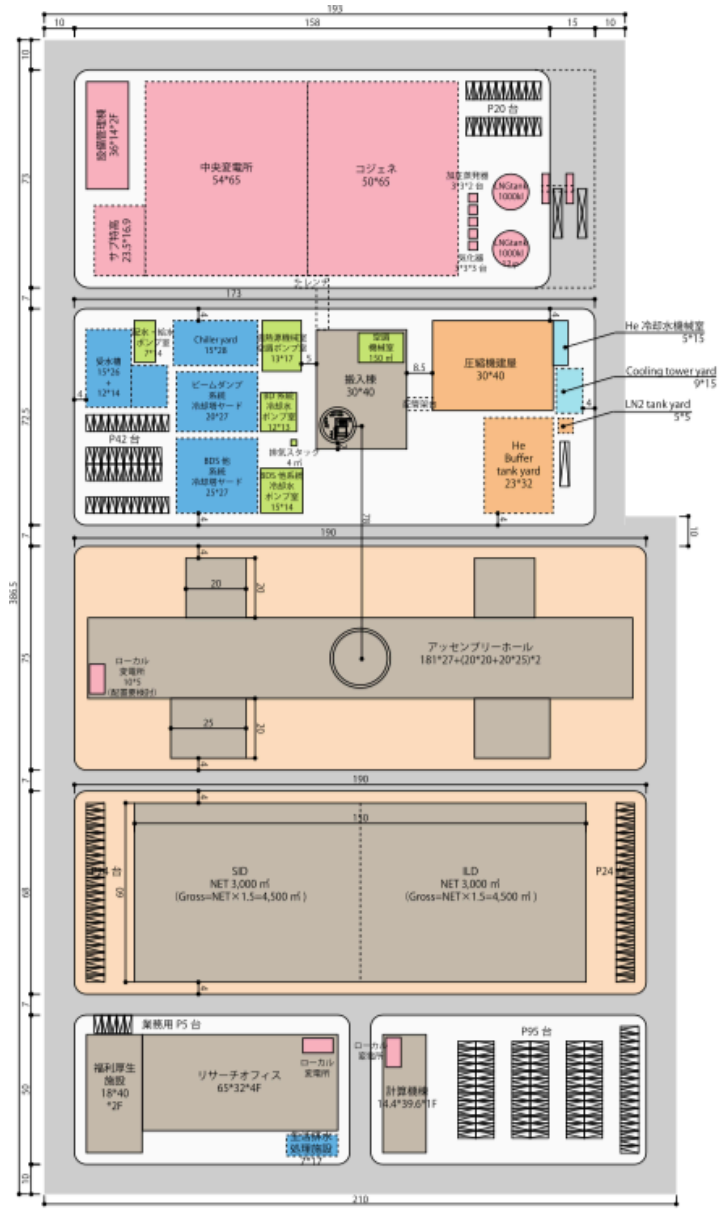

*Figure 5: Functional layout of the interaction region campus shown in Fig. 4* [4].

The time required to accomplish, finally, such a campus is considerable, as can be seen from the Gantt-like chart in Fig. 6. The current assumption is that a preparatory phase of about five to six years will follow a possible 'green light" from the Japanese government to settle the legal procedures, to finalise the land acquisition and to develop the site so that construction work can start. It is assumed that work on the detectors on the site can only start after up to two more years of land development and building construction. Note that in parallel, tunnel construction and the construction of the experimental cavern (DH) is ongoing. These tasks will take considerably longer, so that detector parts cannot be moved underground for a significant time. Figure 7 gives another artistic impression of the overall campus design with all underground tunnels and caverns indicated.

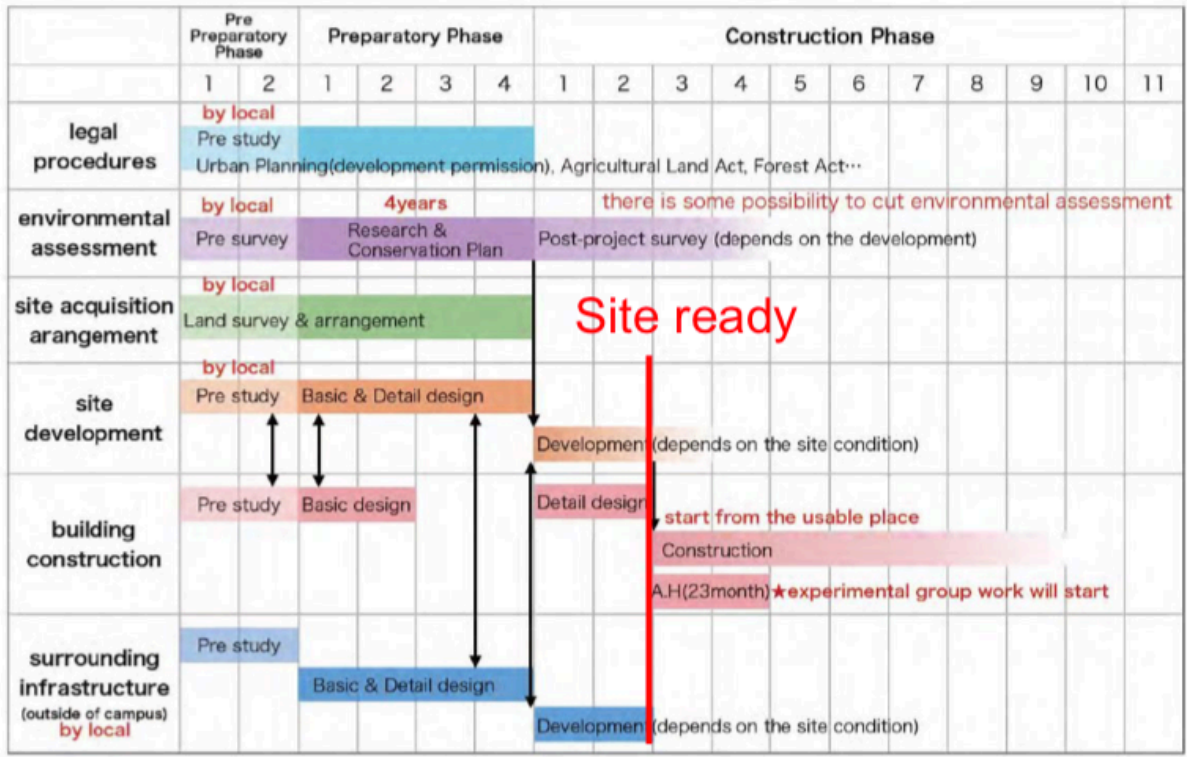

Figure 6: Gantt chart of the time line from green light to the begin of experiment construction [5].

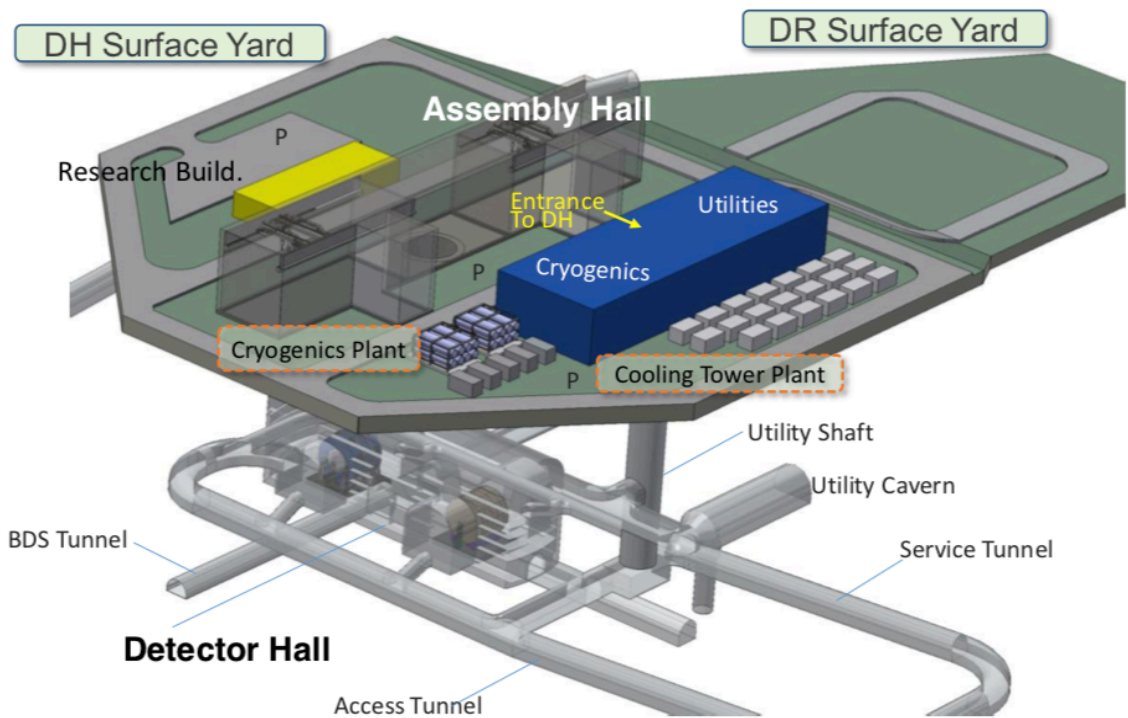

*Figure 7: Artistic impression of the overall interaction region campus, with all over ground and underground facilities indicated. See also Fig. 8 for more details of the underground installations* [6].

## 3. ILD ASSEMBLY PLAN

ILD and SiD are general-purpose high-energy physics detectors. Though different in some conceptional and technological choices, both experiments have a similar outline and will face, to a large extent, similar challenges during construction and assembly. Figure 8 shows the underground areas with the two detectors in "park position", i.e. not in the beam location. Figure 9 is the corresponding technical drawing, this time with one of the two detectors in beam position.

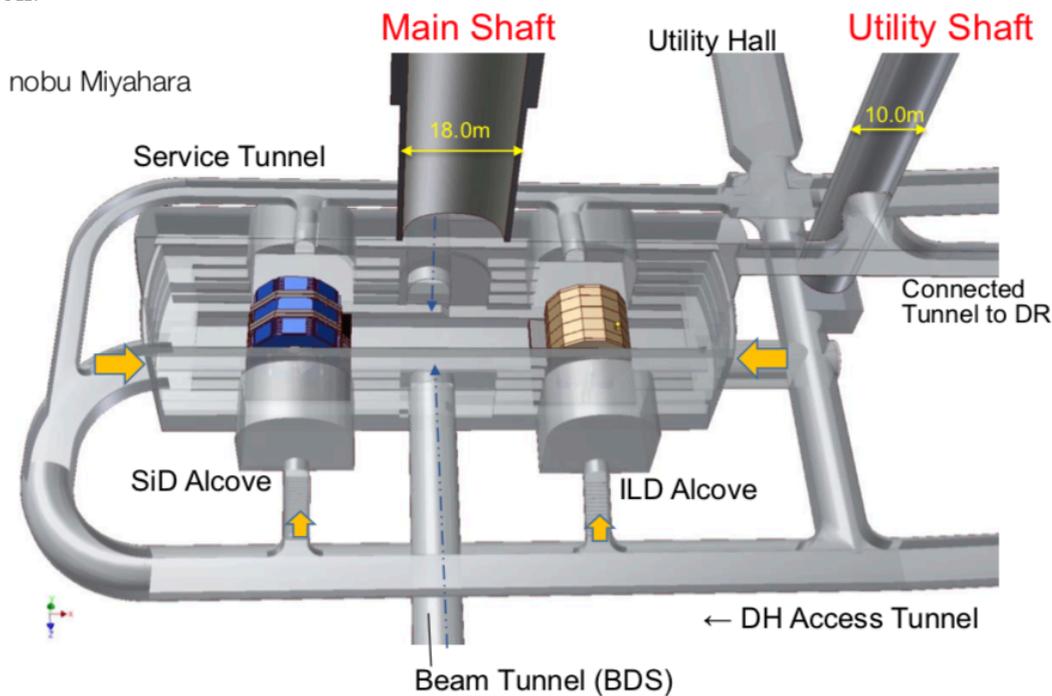

*Figure 8: Overview of underground facilities at the interaction region* [7].

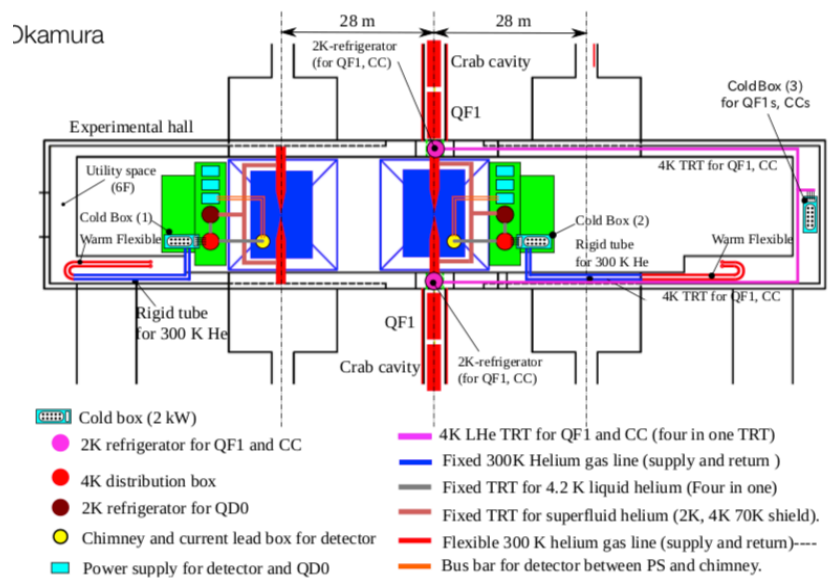

*Figure 9: Technical drawing of the underground detector hall* [8].

An important ingredient to the detector assembly and integration is the main shaft, directly over the detector hall, through which large portions of the ILD and SiD detectors will be lowered. This shaft needs to have large dimensions and requires a heavy-weight gantry crane on top, able to carry, in the case of ILD, the central detector section with cryostat, calorimeters and the yoke ring YB0, with a weight of about 4000t.

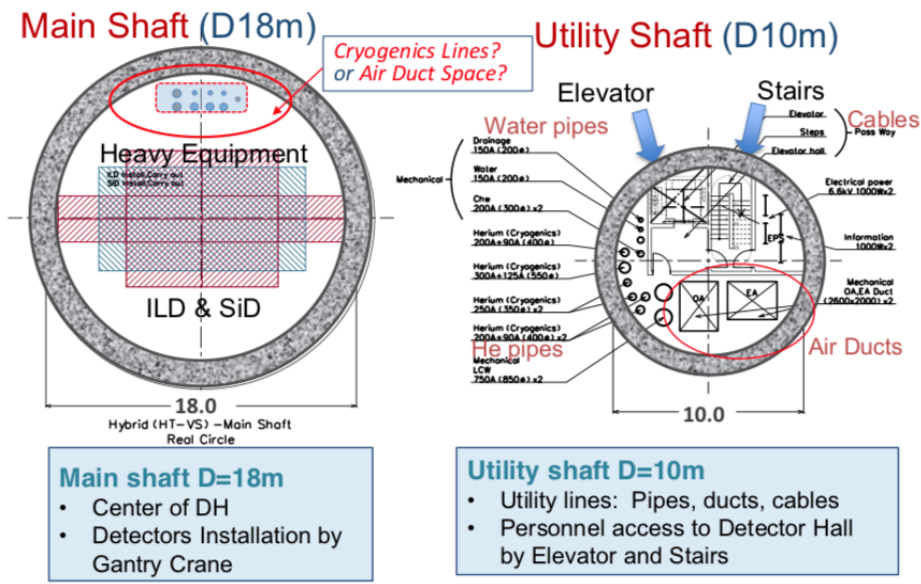

*Figure 10: Cross section of the main shaft and the utility shaft. The main shaft limits the size of individual detector portions that can be lowered to the detector hall* [6].

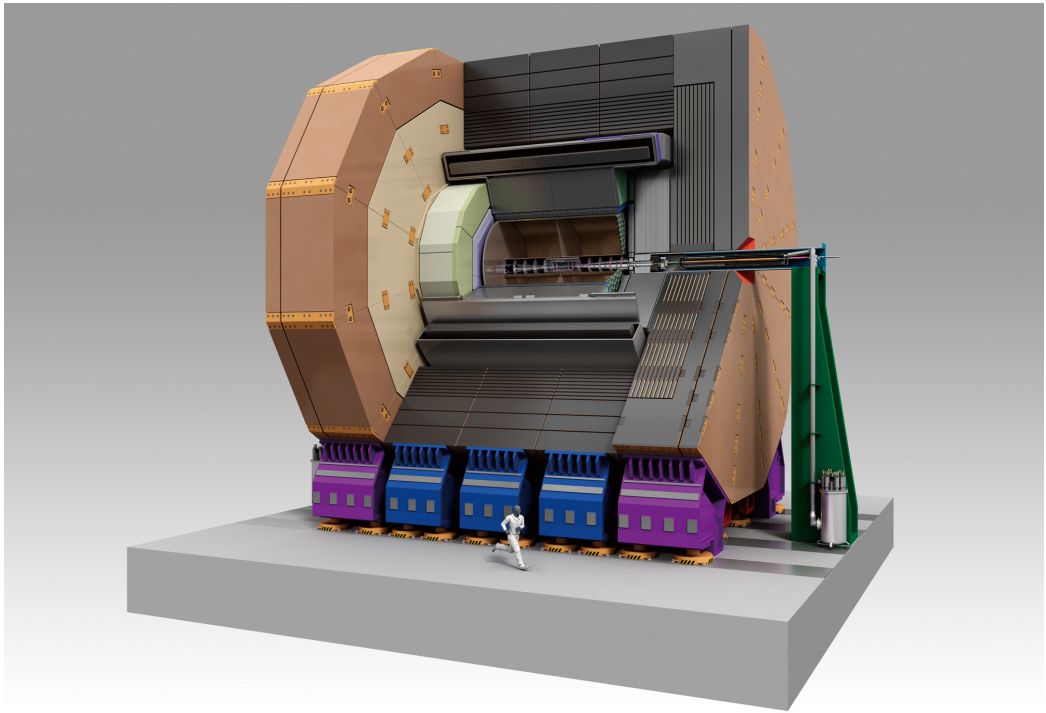

*Figure 11: Technical model of the ILD detector for the ILC [9].*

Both experiments feature the following subdetectors (here listed with their respective challenges concerning construction and assembly):

- Highly precise vertex detectors in silicon technology. These detector components can, in principle be designed and built abywhere in the world, and besides a careful handling and, possibly, clean room installations they have no excessive requirements on the local infrastructures at the interaction region campus.
- Large-volume tracking detectors, either realised in silicon strip technology (SiD) or as a large time projection chamber (TPC, ILD). The SiD silicon can, like the pixel detectors, be constructed anywhere; on site, it has similar requirements (although it is considerably larger). The ILD TPC could either be built elsewhere and the shipped to a nearby harbour and transported by road or rail (provided, e.g., the local roads are sufficiently large), or it might also be constructed on site, in which case it requires significant clean area and some facilities like cranes etc.
- Forward detectors relying on various technologies: these detector parts could be built anywhere and require only limited resources on site.
- Electromagnetic and hadronic calorimeters: The calorimeters belong to the large and heavy detector components and need to be pre-assembled on the campus surface. They require significant storage space, heavy cranes and a careful time planning.
- Instrument iron return yoke: The yokes are heavy steel constructions, with individual parts that challenge logistics, infrastructure and transportation possibilities. Depending on the granularity of the detector planning, individual parts as heavy as more than 200 t need to be transported and integrated into yoke ring structures.
- Solenoid: both experiments feature large (ILD: ~8 m length, ~3.4 m radius), high-field (4 T) solenoids that could either be built at one of the Japanese heavy industries and then transported (in a few large rings) or could be wound on site, requiring large installations. Note that estimates from coil manufacturers assume a solenoid production time of nine years or more, thus, in some scenarios, defining the critical path of the detector assembly.

Figure 11 shows an artist's view of the assembled ILD detector.

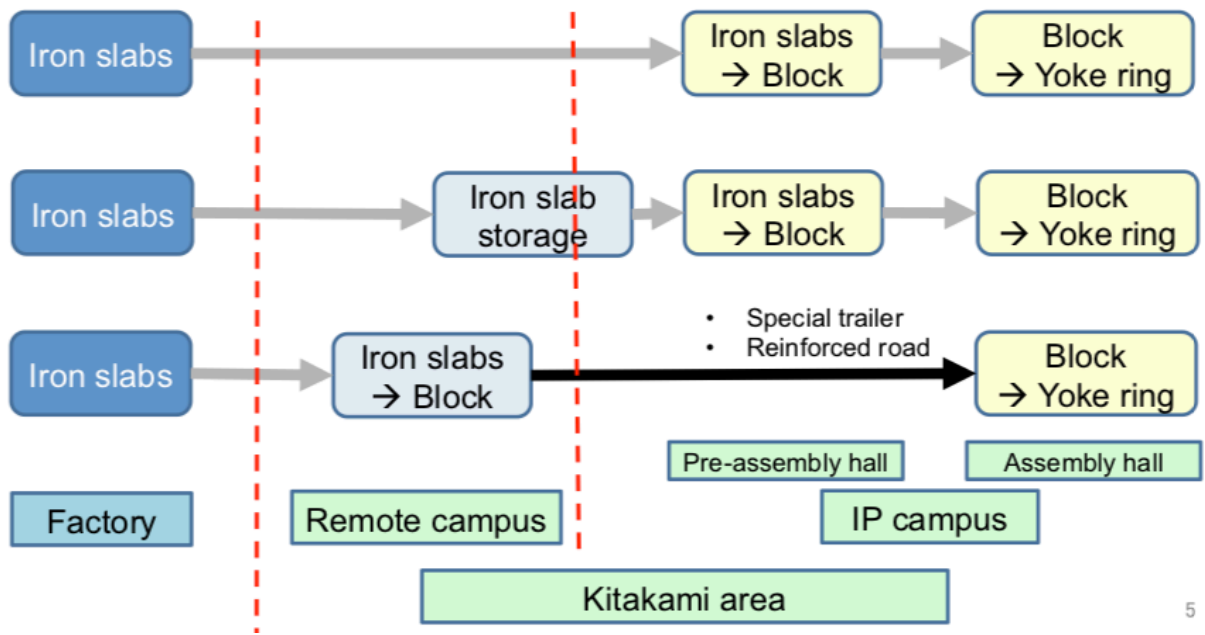

*Figure 12: Various possibilities for assembling the ILD yoke. See text for more details.*

The yoke assembly and the solenoid production nicely demonstrate the significant problems the experiments will face (see also Fig. 12 where individual assembly paths from the individual iron slab to the final yoke block are discussed):
- Size: Transporting large parts – like a third of the ILD solenoid – over rural Kitakami roads might be a challenge due to tunnels, bridges etc. There is also significant risk of damage during transportation. On the other, the effort to produce the solenoids on site is very significant and probably rather expensive.
- Weight: Transportation on Japanese roads is typically limited to 25 t, including the truck itself. This practically precludes the transportation of large yoke parts – which easily reach 90-200 t, depending on the granularity. On the other hand, any production of yoke blocks on site or near the interaction region campus requires large storage space for the iron slabs, the basic elements of the yoke.

- (Storage and assembly) space: Construction or assembly of components on site requires facilities and storage space, and it ties the construction schedule to the availability of these facilities.

Based on these and many similar considerations, assembly plans especially for ILD have been worked out that give a clear picture of the assembly management concerning space (see Fig. 12), time (figs. 13 and 14), and facilities (cranes etc.). These have plans have been discussed at numerous workshops and working meetings, and they will be further scrutinised and refined once green light for ILC construction is given and the detector collaborations have been asked to proceed with their planning[3].

The space in the assembly hall is limited and scarce, and therefore the precise timing of all assembly steps of especially calorimeters, solenoids and yoke need to be carefully planned. Figure 13 shows, as an example, a snapshot of the yoke assembly (green elements) in the ILD half of the AH. The square with two arrows at the top of the figure is is the cover of the main shaft through which the detector portions will have to be lowered.

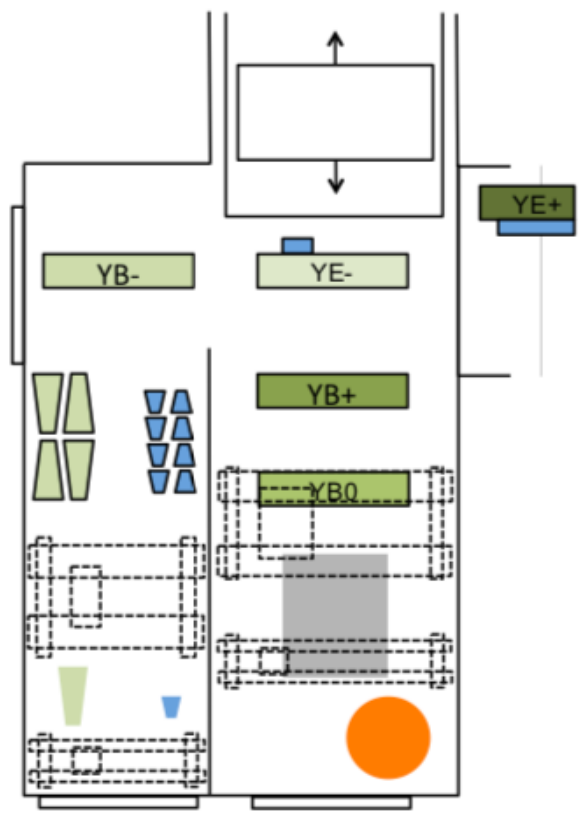

*Figure 13: Snapshot of ILD yoke assembly procedure in the AH.*

---

[3] Presumably once green light is given letters of intent for experiments at the ILC will be asked for from the community, and ILD and SiD are not necessarily the only proposals to be made, nor will their designs remain unchanged until construction begins.

Figure 14 shows, as a further example, a section of the ILD assembly sequence.

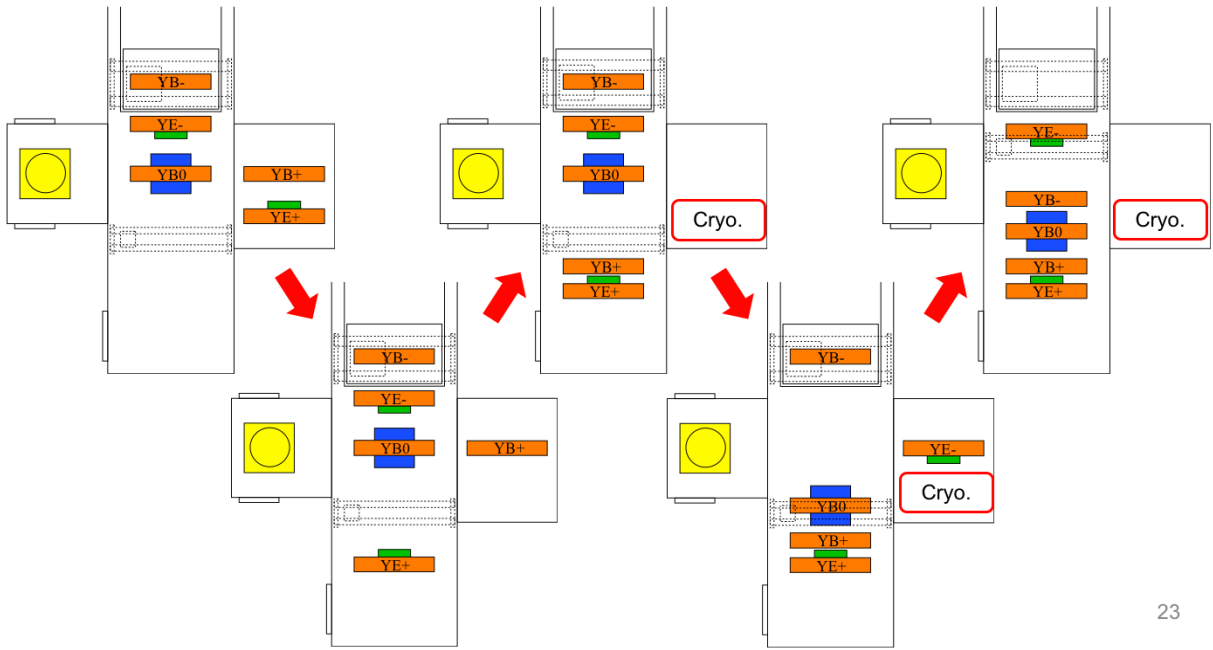

*Figure 14: ILD assembly sequence section.*

A high-level version of the ILD assembly plan is shown in Fig. 15.

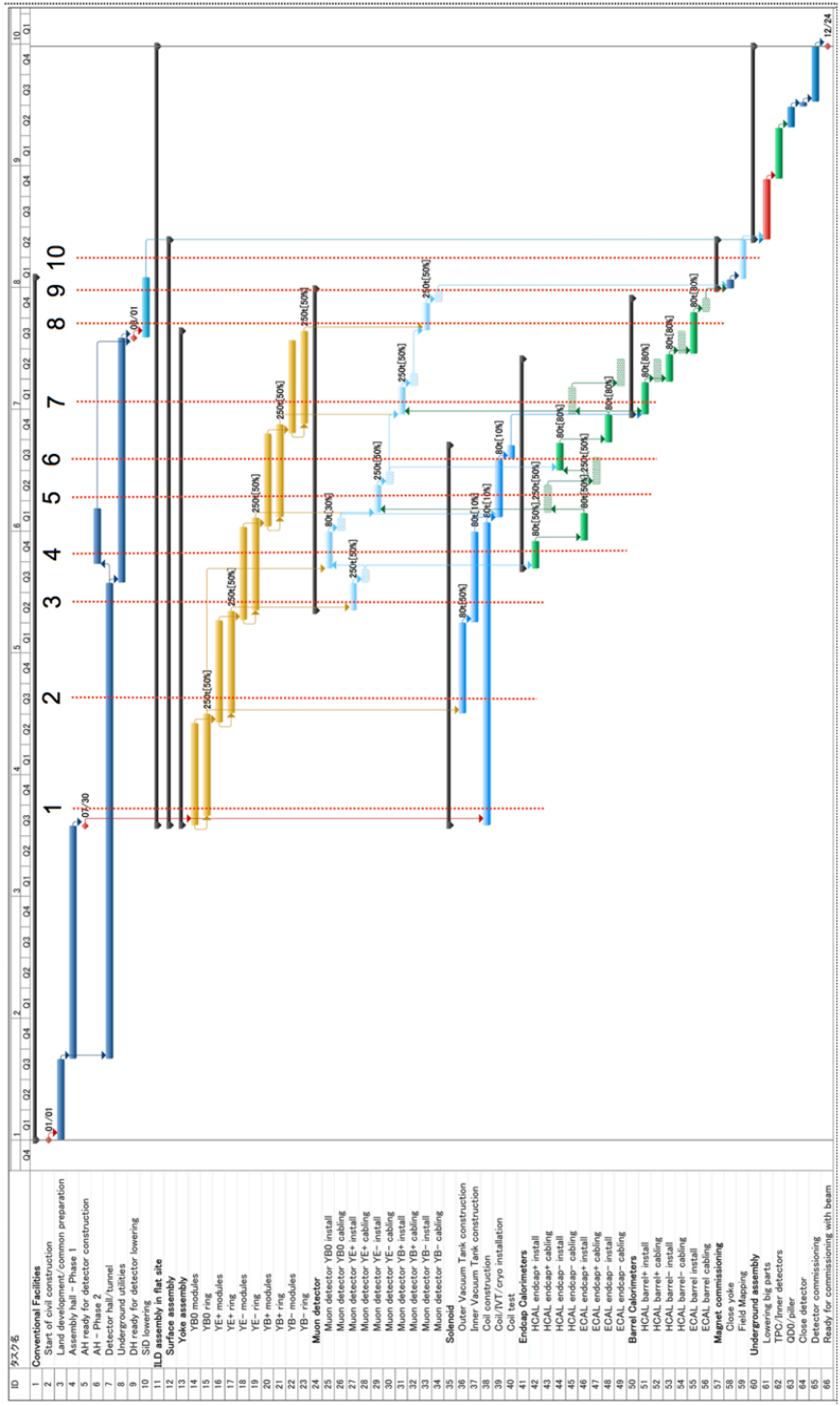

*Figure 15: High-level view of the current ILD assembly plan.*

## 4. CONCLUSIONS

ILD and SiD construction and assembly are involved and complicated procedures that have to respect numerous boundary conditions. Since, however, there is currently no formal ILC project defined, all assembly plans presented here have to be considered as preliminary.

Figures 16 and 17 give two different views of the current understanding of the assembly procedures. While Fig. 16 focuses on the ILD detector components, Fig. 17 includes the time from today over the green light (in year -5) over preparation to the construction.

The important message is, however, that we today have a consistent and concise view of ILD and SiD construction and assembly. E-JADE significantly supported this result through the numerous secondments in work package 3.

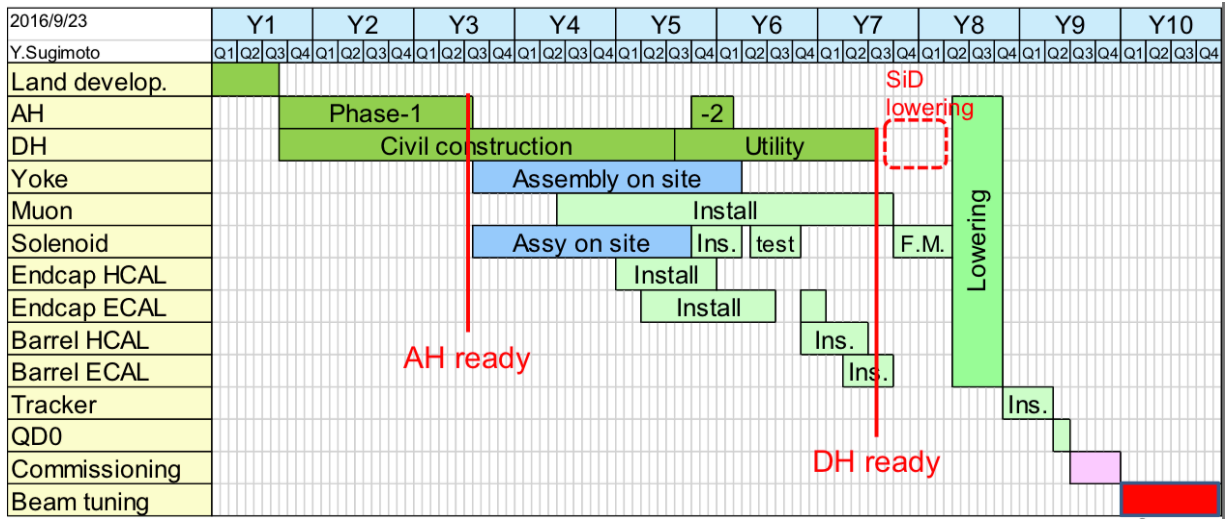

*Figure 16: High-level view of the detector assembly timeline, embedded in the development of the interaction region campus* [10].

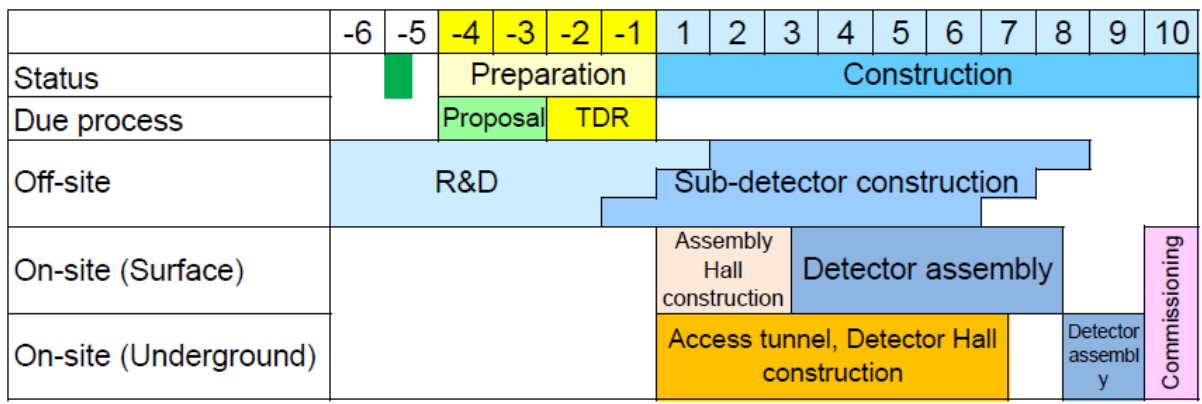

*Figure 17: High-level timeline of the overall ILC project, including the R&D, preparation, and construction phases* [11].

## 5. ILC INFRASTRUCTURE WORKSHOPS

A series workshops on ILC infrastructure for detectors was organised at KEK in Tsukuba, Japan [12]. The agendas are accessible at:

- 30.08.-01.09.2015  -  https://agenda.linearcollider.org/event/6851/
- 14.03.-16.03.2016  -  https://agenda.linearcollider.org/event/6910/
- 30.09.2016  -  https://agenda.linearcollider.org/event/7350/
- 16.05.2017  -  https://agenda.linearcollider.org/event/7611/
- 28.09.-29.09. 2017  -  https://agenda.linearcollider.org/event/7665/
- 23.02.2018  -  https://agenda.linearcollider.org/event/7804/
- 29.11.2018  -  https://agenda.linearcollider.org/event/7976/